\documentclass[12pt]{article}
\usepackage{amssymb}

\input{tcilatex}

\begin{document}

\begin{center}
\textbf{SOLDERED BUNDLE BACKGROUND FOR}

\smallskip\ 

\textbf{THE DE SITTER TOP}

\bigskip\ 

J. Armenta$^{a}$ and J. A. Nieto$^{b}$\footnote[1]{%
nieto@uas.uasnet.mx}

\smallskip\ 

\smallskip\ 

$^{a}$\textit{Departamento de Investigaci\'{o}n en F\'{\i}sica de la
Universidad de Sonora,}

\textit{83000 Hermosillo Sonora, M\'{e}xico}

$^{a}$\textit{Instituto Tecnol\'{o}gico y de Estudios Superiores de}

\textit{Monterrey, Campus Obreg\'{o}n, Apdo. Postal 622, 85000 Cd. Obreg\'{o}%
n Sonora, M\'{e}xico}

$^{b}$\textit{Facultad de Ciencias F\'{\i}sico-Matem\'{a}ticas de la
Universidad Aut\'{o}noma}

\textit{de Sinaloa, 80010 Culiac\'{a}n Sinaloa, M\'{e}xico}

\bigskip\ 

\bigskip\ 

\textbf{Abstract}
\end{center}

We prove that the mathematical framework for the de Sitter top system is the
de Sitter fiber bundle. In this context, the concept of soldering associated
with a fiber bundle plays a central role. We comment on the possibility that
our formalism may be of particular interest in different contexts including
MacDowell-Mansouri theory, two time physics and oriented matroid theory.

\bigskip\ 

\bigskip\ 

\bigskip\ 

\bigskip\ 

\bigskip\ 


Pacs numbers: 04.20.-q, 04.60.+n, 11.15.-q, 11.10.Kk

May, 2005

\newpage

In 1974, Hanson and Regge proposed a Lagrangian theory for a relativistic
top [1]-[2]. One year later it was shown the importance and advantage of
this formulation when the equations of motion of a top in a gravitational
field were derived by using the equivalence gravitational principle [3].
Furthermore, one of the original motivations for the Lagrangian formulation
of the top was to quantize the system by means of the Dirac's method for
constraint Hamiltonian systems. Moreover, with the idea of making
supersymmetric the Lagrangian of the system the square root of a bosonic top
was proposed [4]. In this direction it was shown that the quantum top also
allows a BFV [5] and BRST quantization [6]. It turned out that these pioneer
developments motivated a generalization of the original theory to the
so-called superstringtop theory [7]-[8] which combines, in a Lagrangian
context, the concepts of string and top.

Recently, it has been proposed the de Sitter top theory [9] (see also Ref.
[10]) which is a higher dimensional Lagrangian formulation of a special kind
of a spherical relativistic top characterized by a Regge trajectory
constraint of the de Sitter form $m^{2}+\frac{1}{2r^{2}}\Sigma
^{2}+m_{0}^{2}=0$, where $m$ is the mass of the top, $\Sigma $ is the
internal angular momentum and $r$ and $m_{0}$ are constants. One of the
interesting aspects of the de Sitter top system is that by using the
Kaluza-Klein mechanism [11]-[12] the equation of motion of the top in a
gravitational field can be derived. Since Kaluza-Klein theory is closely
linked to the fiber bundle concept [13]-[14] one should expect a geometric
formulation of the de Sitter top in terms of such a concept. Although this
idea has been outlined in Refs. [9] and [10] the precise connection between
fiber bundle structure and the de Sitter top needs to be addressed. In this
work, we claim that the soldered fiber bundle structure is the natural
mathematical framework for a formal description of the de Sitter top. It
turns out that the soldered bundle concept has been extensively used by
Drechsler (see Ref. [15] and references therein) in particle physics. In
particular, Drechsler has proposed a gauge theory for extended elementary
objects based in the soldered bundle concept. These applications of the
soldered bundle concept are, however, more focused on a gauge field scenario
of the systems [16]-[17] rather than in a gravitational context. Moreover,
part of the motivation of these applications are at the level of hadrons
[18]-[21] and not a deeper level as is the case of the de Sitter top.
Nevertheless, in this work we show that some of the mathematical tools in
the soldered bundle theory can be used for describing the de Sitter top. We
complement our analysis of the de Sitter top by observing that the soldered
bundle concept may provide the mathematical tool to clarify some aspects of
the MacDowell-Mansouri formalism [22] (see also Ref. [23]) and two time
physics [24].

Let us start writing the de Sitter top Lagrangian [9]

\begin{equation}
L=-m_{0}(-\frac{1}{2}\omega _{M}^{AB}\omega _{NAB}\dot{x}^{M}\dot{x}%
^{N})^{1/2},  \tag{1}
\end{equation}%
where $m_{0}$ is a constant measuring the inertia of the system and $\dot{x}%
^{M}=\frac{d}{d\tau }x^{M}$, with $M,N$ running from $0$ to $9$ and $\tau $
is an arbitrary parameter. Here, $\omega _{M}^{AB}=-\omega _{M}^{BA}$
denotes a connection associated with the de Sitter group $SO(1,4)$ (or anti
de Sitter group $SO(2,3)$). Indeed, $\omega _{M}^{AB}$ is determined by the
ansatz

\begin{equation}
\omega _{M}^{AB}=\left( 
\begin{array}{cc}
\omega _{\mu }^{5a}\left( x\right) & \omega _{\mu }^{ab}\left( x\right) \\ 
0 & \omega _{i}^{ab}\left( y\right)%
\end{array}%
\right) .  \tag{2}
\end{equation}%
When writing (2) the coordinates $x^{M}$ were separated in the form $(x,y),$
with $x$ corresponding to the four dimensional base manifold $M^{4}$ and $y$
parametrizing the group manifold $SO(1,4)$. The Lagrangian (1) is
interesting because it leads to a Regge trajectory constraint of the de
Sitter form $m^{2}+\frac{1}{2r^{2}}\Sigma ^{2}+m_{0}^{2}=0$ (see Refs.
[1]-[2] and also Refs. [25]-[26]) connecting the mass $m$ and the spin $%
\Sigma $ of the system.

Consider the antisymmetric pair $[ab]$ of the four valued indices $a,b$ in
the form $a^{\prime }=([12],[13],[14],[23],[24],[34])$. Using this notation
one discovers that if one makes the identifications ($\sim $),%
\begin{equation}
\omega _{\mu }^{5a}\left( x\right) \sim e_{\mu }^{a}\left( x\right) , 
\tag{3}
\end{equation}

\begin{equation}
\omega _{\mu }^{ab}\left( x\right) \sim E_{\mu }^{a^{\prime }}\left(
x\right) ,  \tag{4}
\end{equation}

\begin{equation}
\omega _{i}^{ab}\left( y\right) \sim E_{i}^{a^{\prime }}\left( y\right) 
\tag{5}
\end{equation}%
and%
\begin{equation}
\omega _{M}^{AB}\rightarrow E_{M}^{\hat{A}}  \tag{6}
\end{equation}%
then the ansatz (2) becomes

\begin{equation}
E_{M}^{\hat{A}}=\left( 
\begin{array}{cc}
e_{\mu }^{a}\left( x\right) & E_{\mu }^{a^{\prime }}\left( x\right) \\ 
0 & E_{i}^{a^{\prime }}\left( y\right)%
\end{array}%
\right) ,  \tag{7}
\end{equation}%
which may be recognized as the typical form of the Kaluza-Klein ansatz
[11]-[12]. This suggests to introduce the metric

\begin{equation}
\gamma _{MN}=E_{M}^{\hat{A}}E_{N}^{\hat{B}}\eta _{\hat{A}\hat{B}},  \tag{8}
\end{equation}%
which according to (7) can be separated in the form

\begin{equation}
\begin{array}{l}
\gamma _{\mu \nu }=g_{\mu \nu }+E_{\mu }^{a^{\prime }}E_{\nu }^{b^{\prime
}}\eta _{a^{\prime }b^{\prime }}, \\ 
\\ 
\gamma _{\mu i}=E_{\mu }^{a^{\prime }}E_{i}^{b^{\prime }}\eta _{a^{\prime
}b^{\prime }}, \\ 
\\ 
\gamma _{ij}=E_{i}^{a^{\prime }}E_{j}^{b^{\prime }}\eta _{a^{\prime
}b^{\prime }}=g_{ij}(y),%
\end{array}
\tag{9}
\end{equation}%
where%
\begin{equation}
g_{\mu \nu }(x)=e_{\mu }^{a}\left( x\right) e_{\nu }^{b}\left( x\right) \eta
_{ab}.  \tag{10}
\end{equation}%
Here, $\eta _{ab}$ and $\eta _{a^{\prime }b^{\prime }}$ are flat metrics
corresponding to $M^{4}$ and $B$ respectively.

From of the point of view of Kaluza-Klein theory the splitting (9) is the
result of compactifying a $4+D-$dimensional space-time manifold $M^{4+D}$ in
the form $M^{4+D}\rightarrow M^{4}\times B$, where $M^{4}$ is a
four-dimensional base space manifold and $B$ is a group manifold whose
dimension is $D$. In the case of the de Sitter top it is not clear what it
the meaning of $M^{4+D}$ and $B$ is. Moreover, the meaning of the
identification (3) is unclear. One may choose $M^{4+D}$ as $SO(1,4)$ and $B$
as $SO(1,3)$ but in this case $M^{4}$ will be completely determined without
using the gravitational field equations. Thus, although the identifications
(3)-(6) are suggestive they are not complete requiring a deeper analysis. We
shall show that these aspects of the de Sitter top can be clarified through
the so-called \textit{soldered fiber bundle} notion. In order to achieve our
goal we first observe that the object $\omega _{M}^{AB}$ may be identified
with the fundamental $1$-form

\begin{equation}
\omega =g^{-1}dg+g^{-1}\Omega g,  \tag{11}
\end{equation}%
where

\begin{equation}
\omega =\frac{1}{2}\omega _{M}^{AB}S_{AB}dx^{M}  \tag{12}
\end{equation}%
and

\begin{equation}
\Omega =\frac{1}{2}\omega _{\mu }^{AB}S_{AB}dx^{\mu }.  \tag{13}
\end{equation}%
Here, $g\in SO(1,4)$ and $S_{AB}$ are the generators of the de Sitter group $%
SO(1,4)$ (or anti de Sitter group $SO(2,3)$). In turn, (13) can be
understood as a $1$-form connection in the cotangent space $T^{\ast }(P)$,
where $P$ is a principal bundle $P(M^{4},SO(1,4))$. Locally, $%
P(M^{4},SO(1,4))$ looks like $M^{4}\times SO(1,4)$ but once again this
picture is incomplete in the sense that it leaves without answer the meaning
of the relation (3). Nevertheless, this analysis motivates to find a
framework beyond the simple principal bundle $P(M^{4},SO(1,4))$.

In general, it is well known that given a principal bundle $P(M,G)$ one may
construct an associated fiber bundle $E(M,F,G)$ where $F$ is a fiber
manifold and conversely, a fiber bundle $E(M,F,G)$ naturally induces a
principal bundle $P(M,G)$ associated with it (see Ref. [27], section 9.4.2,
for details). Thus, the principal bundle $P(M^{4},SO(1,4))$ may have an
associated fiber bundle $E(M^{4},F,SO(1,4))$. In principle $F$ can be any
vector space but in the case of the de Sitter top one may think of $SO(1,4)$
as a group acting transitively over $F$. Moreover, by In\"{o}n\"{u}-Wigner
contraction one should expect that de Sitter top is reduced to the usual
relativistic top which is invariant under the Lorentz transformations
determined by the Lorentz group $SO(1,3)$ (see Ref. [2]). Now, since $%
SO(1,3) $ is an isotropy subgroup of $SO(1,4)$ this suggest to consider the
coset space $SO(1,4)/SO(1,3)$. Applying a well-known theorem (see Ref. [28],
section 1.6, and see also Ref. [29]) one can establish the homeomorphism $%
F\cong SO(1,4)/SO(1,3).$ Therefore, the fiber bundle which we are interested
to relate to the de Sitter top is

\begin{equation}
E(M^{4},F\cong SO(1,4)/SO(1,3),SO(1,4)).  \tag{14}
\end{equation}%
This is a fiber bundle of the Cartan type possessing as a fiber de Sitter
space $dS_{4}\equiv F$ which is homeomorphic to the noncompact coset space 
\[
SO(1,4)/SO(1,3). 
\]
Since $\dim G/H=\dim G-\dim H$ we find that%
\[
\dim SO(1,4)/SO(1,3)=\dim SO(1,4)-\dim SO(1,3)=4 
\]%
and therefore the dimension of $dS_{4}$ is four, the same as $M^{4}$. This
result is an indication that the bundle (14) may admit a soldered fiber
bundle structure. But before we present the definition of a soldered fiber
bundle let us motivate further the idea of soldering in connection with the
de Sitter top.

Let us introduce a gauge parameter $\lambda =\frac{1}{2}\lambda
^{AB}(x)S_{AB}.$ The transformation associates with $\omega ,$ given in
(12), can be written as

\begin{equation}
\omega ^{\prime }=g\omega g^{-1}+g^{-1}dg,  \tag{15}
\end{equation}%
which as an infinitesimal gauge transformation reads as

\begin{equation}
\delta \omega =d\lambda +[\lambda ,\omega ].  \tag{16}
\end{equation}%
In components (16) leads to the expression

\begin{equation}
\delta \omega ^{AB}=d\lambda ^{AB}+\omega ^{AC}\lambda _{C}^{B}+\omega
^{AC}\lambda _{C}^{B},  \tag{17}
\end{equation}%
which can be separated in the form%
\begin{equation}
\delta \omega ^{5a}=D\lambda ^{5a}+\omega ^{5c}\lambda _{c}^{a}  \tag{18}
\end{equation}%
and

\begin{equation}
\delta \omega ^{ab}=D\lambda ^{ab}+\omega ^{a5}\lambda _{5}^{b}+\omega
^{b5}\lambda _{5}^{a}.  \tag{19}
\end{equation}%
Here, $D$ means covariant derivative with $\omega ^{ab}$ as a connection.
Thus, if we set

\begin{equation}
\omega _{\mu }^{5a}=e_{\mu }^{a}  \tag{20}
\end{equation}%
one sees that the expressions (18) and (19) lead to

\begin{equation}
\delta e_{\mu }^{a}=D_{\mu }\xi ^{a}+e_{\mu }^{c}\lambda _{c}^{a}  \tag{21}
\end{equation}%
and

\begin{equation}
\delta \omega _{\mu }^{ab}=D_{\mu }\lambda ^{ab}+e_{\mu }^{a}\xi ^{b}-e_{\mu
}^{b}\xi ^{a},  \tag{22}
\end{equation}%
respectively. But the formulae (21) and (22) indicate that neither $e_{\mu
}^{a}$ nor $\omega _{\mu }^{ab}$ transform properly under Lorentz
transformations $SO(1,3)$ and therefore in general they cannot be identified
with the Lorentz tetrad and connection respectively. For this to be possible
it is required that the parameter $\xi ^{a}$ vanishes in (21) and (22). In
turn this implies that the de Sitter group $SO(1,4)$ is broken leading to
the Lorentz group $SO(1,3)$. To set the parameter $\xi ^{a}$ equal to zero
in a consistent way is not so simple and in fact requires to introduce the
soldering concept which we shall now formally define.

A bundle $E(M,F=G/H,G)$ over a base manifold $B$ with homogeneous fiber $%
F=G/H$ and associated principle bundle $P(M,G)$ is soldered if [15]

(i) $G$ acts transitively on $F$,

(ii) $\dim F=\dim M$

(iii) $E$ admits a global cross section $\sigma $ and the structural group $%
G $ of $E$ can be reduced to $H$.

(iv) The tangent bundle $T(M)$ over $M$ and the bundle $T(E)=T(F)$ of all
tangent vectors to $F$ at the section $\sigma $ are isomorphic.

It is clear that the bundle $E(M^{4},F\cong SO(1,4)/SO(1,3),SO(1,4))$ which
we try to associate to the de Sitter top satisfies (i), (ii) and (iii). The
only remaining point is to impose the condition (iv). It turns out that (iv)
is equivalent to the two properties [15]

$(a)$ $\theta (X^{\prime })=0$ for $X^{\prime }\in T(P^{\prime }(M,H))$ if
and only if $X^{\prime }$ is vertical.

$(b)$ $\theta (dR_{h}X^{\prime })=h^{-1}X^{\prime }h$ for $X^{\prime }\in
T(P^{\prime }(M,H))$ and $h\in H,$

\noindent which are satisfied for the so-called $1$-form of soldering $%
\theta $. The reason to be interested in the $1$-form $\theta $ is because a
connection $\omega $ in $P(M,G)$ can be written in terms of a connection in $%
P^{\prime }(M,H)$ and the soldering $1$-form $\theta $ as follows

\begin{equation}
\omega =\omega ^{\prime }+\theta .  \tag{23}
\end{equation}%
Actually, the equivalence between (iv) and $(a)-(b)$ is achieved when one
assume the reductive algebra

\begin{equation}
\lbrack \mathcal{L}(H),\Lambda ]\subseteq \Lambda ,  \tag{24}
\end{equation}%
associated with the decomposition $\mathcal{L}(G)=\mathcal{L}(H)\oplus
\Lambda $ of the Lie algebra of $G$, where $\mathcal{L}(H)$ corresponds to
the subalgebra of $H$ and $\Lambda $ is required to be a vector subspace of $%
G$ with dimension $\dim M=\dim G-\dim H$. In fact such equivalence is
obtained when one assumes the existence of $\Lambda $\textit{-}valued $1$%
-form $\theta $ satisfying $(a)-(b)$ (see Refs. [15] and [42] for details).
In connection with (23) two observations are necessary. First, the
decomposition (23) do not require the additional condition

\begin{equation}
\lbrack \Lambda ,\Lambda ]\subseteq \mathcal{L}(H),  \tag{25}
\end{equation}%
implying that $F\cong SO(1,4)/SO(1,3)$ is a symmetric space, and second,
since $\omega $ is a $\mathcal{L}(G)$\textit{-}valued $1$-form in the
cotangent bundle $T^{\ast }(P)$ we can write $\omega =\omega (x,y)$, where
the coordinates $x$ and $y$ parametrize locally $M$ and $F$ respectively,
and consequently, in general, we should have $\omega ^{\prime }=\omega
^{\prime }(x,y)$ and $\theta =\theta (x,y)$. However, $(a)$ and $(b)$
implies that we can write $\theta =\theta (x).$

Let us now specialize (23) to the case of the de Sitter fiber bundle $%
E(M^{4},F\cong SO(1,4)/SO(1,3),SO(1,4)).$ First let us observe that since $%
S_{AB}$ are the generators of the de Sitter group $SO(1,4)$ (or anti de
Sitter group $SO(2,3)$) we can write the algebra

\begin{equation}
-i[S_{AB},S_{CD}]=\eta _{AC}S_{BD}-\eta _{AD}S_{BC}+\eta _{BD}S_{AC}-\eta
_{BC}S_{AD},  \tag{26}
\end{equation}%
where $\eta _{AC}=diag(-1,1,1,1,1).$ From (26) we get

\begin{equation}
-i[S_{ab},S_{cd}]=\eta _{ac}S_{bd}-\eta _{ad}S_{bc}+\eta _{bd}S_{ac}-\eta
_{bc}S_{ad},  \tag{27}
\end{equation}%
\begin{equation}
-i[S_{5b},S_{cd}]=\eta _{bd}S_{5c}-\eta _{bc}S_{5d},  \tag{28}
\end{equation}%
and%
\begin{equation}
-i[S_{5b},S_{5d}]=S_{bd}.  \tag{29}
\end{equation}%
Thus, we conclude that $S_{ab}$ are the generators of the subgroup $SO(1,3)$
and that (28) is in agreement with (24). This in turn implies that the $%
SO(1,4)$ valued $1$-form connection $\omega $ given in (12) can be written
in terms of a $SO(1,3)$ valued $1$-form connection $\omega ^{ab}$ and the $1$%
-form $\omega ^{5b}$ as

\begin{equation}
\omega =\frac{1}{2}\omega ^{ab}S_{ab}+\omega ^{5b}S_{5b}.  \tag{30}
\end{equation}%
Comparing (30) with (23) one discovers that one can make the identifications 
$\omega ^{\prime ab}=\omega ^{ab}$ and $\theta ^{b}=\omega ^{5b}$. Since in
general $\omega ^{AB}=\omega ^{AB}(x,y)$ (see expressions (11)-(12)), one
finds that $\omega ^{AB}$ can be split into the form%
\begin{equation}
\theta ^{a}=\omega _{\mu }^{5a}dx^{\mu }+\omega _{i}^{5a}dy^{i}  \tag{31}
\end{equation}%
and

\begin{equation}
\omega ^{ab}=\omega _{\mu }^{ab}dx^{\mu }+\omega _{i}^{ab}dy^{i}.  \tag{32}
\end{equation}%
Now, the condition $(a)-(b)$ for $\theta $ means that the soldering concept
allows to set $\omega _{i}^{5a}=0$ and therefore we have

\begin{equation}
\theta ^{a}=\omega _{\mu }^{5a}dx^{\mu }.  \tag{33}
\end{equation}%
This is consistent with (iv) and in fact one should be able to write $\theta
^{a}$ in terms of the base $e_{\mu }^{a}\in T(M^{4})$. In particular one can
set $\theta _{\mu }^{a}=e_{\mu }^{a}$, that is, $\omega _{\mu }^{5a}=e_{\mu
}^{a}$. Consequently, one discovers that the complete connection $\omega
_{M}^{AB}$ can be written as

\begin{equation}
\omega _{M}^{AB}=\left( 
\begin{array}{cc}
e_{\mu }^{a}\left( x\right) & \omega _{\mu }^{ab}\left( x,y\right) \\ 
0 & \omega _{i}^{ab}\left( x,y\right)%
\end{array}%
\right) ,  \tag{34}
\end{equation}%
which by using the Kaluza-Klein mechanism can be reduce to (2). Summarizing,
we have explicitly shown that the geometrical framework for the de Sitter
top is the de Sitter soldering fiber bundle $E(M^{4},F\cong
SO(1,4)/SO(1,3),SO(1,4))$ as Fukuyama had anticipated [9].

Until now we have focused more on the de Sitter top description determined
by the line element (1) rather than in the dynamics of the background itself
where the system moves. In other words, besides the mathematical framework
for the de Sitter top one may be interested in the field equations which
govern the evolution of the connection $\omega _{M}^{AB}$ given in (2).
Since $e_{\mu }^{a}$ and $\omega _{M}^{ab}\left( x\right) $ are considered
as independent variables one may think in Einstein-Cartan theory (linear in
the curvature), as the more appropriate candidate. However, there exist
another gravitational theory which seems to be closer to the spirit of the
ansatz (2) than the Einstein-Cartan theory. We refer to the so-called
MacDowell-Mansouri theory [22] (see also Ref. [23]) which is one of the
closest proposals for achieving a gauge theory for gravity. The idea in this
theory is precisely to consider the field variables $e_{\mu }^{a}$ and $%
\omega _{\mu }^{ab}\left( x\right) $ as part of a bigger connection $\omega
_{\mu }^{AB}\left( x\right) $ associated with the de Sitter group $SO(1,4)$.
In fact, by taking $e_{\mu }^{a}=$ $\omega _{\mu }^{5a}\left( x\right) $ one
verifies that the action

\begin{equation}
S=\int d^{4}x\varepsilon ^{\mu \nu \alpha \beta }\mathcal{R}_{\mu \nu }^{ab}%
\mathcal{R}_{\alpha \beta }^{cd}\varepsilon _{abcd},  \tag{35}
\end{equation}%
where

\begin{equation}
\mathcal{R}_{\mu \nu }^{ab}=R_{\mu \nu }^{ab}+e_{\mu }^{a}e_{\nu
}^{b}-e_{\mu }^{b}e_{\nu }^{a},  \tag{36}
\end{equation}%
with $R_{\mu \nu }^{ab}$ the curvature in terms of $\omega _{\mu }^{ab}$,
leads to the two terms: the Euler-Pontrjagin topological invariant and the
Einstein Hilbert action with a cosmological constant. One observes that the
action (35) is intrinsically four dimensional and therefore there seems to
be enough room for the additional part $\omega _{i}^{ab}\left( y\right) $.
However, recently [30] in an effort to generalize the Ashtekar formalism to
higher dimensions it has been proposed a generalization of (35) to eight
dimensions which may allow full background description of the connection
(2). In fact, in reference [30] it was proposed the action

\begin{equation}
S=\int d^{4}x\eta ^{MNRS}\mathcal{R}_{MN}^{\hat{A}\hat{B}}\mathcal{R}_{MN}^{%
\hat{C}\hat{D}}\eta _{\hat{A}\hat{B}\hat{C}\hat{D}},  \tag{37}
\end{equation}%
where the object $\eta ^{MNRS}$ is connected with the algebra of octonions
(see Ref. [30] for details). Our conjecture is that the background field
equations, where the de Sitter top evolves, can be derived from the action
(37). Assuming that the action (37) describes the evolution of the de Sitter
connection $\omega _{M}^{\hat{A}\hat{B}}$ we observe that an interesting
possibility arises. This has to do with the fact that a theory based on the
action (30) would lead to a de Sitter vacuum solution for the base manifold $%
M^{4}$ rather than the Minkowski space. Consequently, the de Sitter
soldering fiber bundle $E(M^{4},F\cong SO(1,4)/SO(1,3),SO(1,4))$ will be
such that both $M^{4}$ and the fiber $F$ are de Sitter (or anti-de Sitter)
type spaces. Besides these observations our formalism may help to understand
why the choice $e_{\mu }^{a}=$ $\omega _{\mu }^{5a}\left( x\right) $ works
in this context. In general this has been a mystery, but according to our
discussion such a choice is the result of a soldering process associated
with the fiber bundle 
\[
E(M^{4},F\cong SO(1,4)/SO(1,3),SO(1,4)). 
\]

Our observation that the de Sitter top may be described by the de Sitter
soldering bundle may also be useful for a possible connection between the de
Sitter top and two time physics. In fact, it turns out that two time physics
is determined by the action [24]%
\begin{equation}
S=\int_{\tau _{i}}^{\tau _{f}}d\tau \left( \frac{1}{2}\varepsilon ^{ab}\dot{x%
}_{a}^{\mu }x_{b}^{\nu }\eta _{\mu \nu }-\frac{1}{2}\lambda ^{ab}\left(
x_{a}^{\mu }x_{b}^{\nu }\eta _{\mu \nu }+m_{ab}^{2}\right) \right) , 
\tag{38}
\end{equation}%
where $x_{1}^{\mu }=x_{a}^{\mu },$ $x_{2}^{\mu }=p^{\mu },\lambda
^{ab}=\lambda ^{ba}$ is a Lagrange multiplier and $m_{ab}^{2}$ are constants
which can be zero or different from zero. If $m_{ab}^{2}=0$ then the action
(38) has the manifest $Sp(2,R)$ (or $SL(2,R))$ invariance and the action is
consistent if the flat metric $\eta _{\mu \nu }$ admit signature with two
time. However if $m_{ab}^{2}\neq 0$ this symmetry is broken and the action
leads to the constraint

\begin{equation}
\Omega _{ab}=x_{a}^{\mu }x_{b}^{\nu }\eta _{\mu \nu }+m_{ab}^{2}=0  \tag{39}
\end{equation}%
(see Ref. [31] for details). Choosing $m_{11}^{2}=-R^{2}$, $%
m_{22}^{2}=m_{0}^{2}$ and $m_{12}^{2}=0$ one discovers that (39) gives

\begin{equation}
x^{\mu }x_{\mu }-R^{2}=0,  \tag{40}
\end{equation}

\begin{equation}
x^{\mu }p_{\mu }=0  \tag{41}
\end{equation}%
and

\begin{equation}
p^{\mu }p_{\mu }+m_{0}^{2}=0.  \tag{42}
\end{equation}%
The formula (40) describes an anti-de Sitter spacetime and therefore in a
sense the system can be understood as a relativistic point particle moving
in a anti-de Sitter background which is precisely the idea underlying the de
Sitter top.

Finally, there are at least two possible interesting generalizations of our
formalism for the de Sitter top. In the first case one may think in the de
Sitter top as a result of Clifford geometry as presented by Castro (see Ref.
[32] and references therein). The second possibility may arise from the
so-called oriented matroid theory [33]. It has been shown that oriented
matroid theory can be connected not only to string theory but also to any $%
p- $brane, supergravtiy and Chern-Simons theory [34]-[38]. These connections
were possible thanks to the notion of \textit{matroid bundle} introduced
first by MacPherson [39] and generalized by Anderson and Davis [40] and Biss
[41]. It turns out that matroid bundle is a generalization of the concept of
a fiber bundle. Thus, it seems natural to associate with the bundle $%
E(M^{4},F\cong SO(1,4)/SO(1,3),SO(1,4))$ some kind of the de Sitter matroid
bundle and therefore a de Sitter matroid \ top or a de Sitter "topoid". At
present all these possibilities concerning the background (2) are under
investigation and we expect to report our results somewhere in the not too
distant future.

\bigskip\ 

\noindent \textbf{Acknowledgments: }J. A. Nieto would like to thank E.
Sezgin and Texas A\&M University Physics Department for their kind
hospitality, where part of this work was developed.

\end{document}